\documentclass[prd,aps,twocolumn]{revtex4}
\usepackage{bm}
\usepackage{graphicx} 
\usepackage{amsmath,amssymb}
\usepackage{aas_macros}
\usepackage{color}
\usepackage{ulem}

\begin{document}
\date{\today} 
\title{Upper limit on the amplitude of gravitational waves 
around 0.1Hz from the Global Positioning System} 

\author{Shohei Aoyama}
\affiliation{Department of Physics and Astrophysics, Nagoya University,
Nagoya 464-8602, Japan} 

\author{Rina Tazai} 
\affiliation{Department of
Physics and Astrophysics, Nagoya University, Nagoya 464-8602, Japan}

\author{Kiyotomo Ichiki} 
\affiliation{Kobayashi-Maskawa Institute for
the Origin of Particles and the Universe, Nagoya University, Nagoya
464-8602, Japan}

\begin{abstract}
The global positioning system (GPS) is composed of thirty one satellites
having atomic clocks with $10^{-15}$ accuracy on board and enables
one to calibrate the primary standard for frequency on the ground.
Using the fact that oscillators on the ground have been successfully
stabilized with high accuracy by receiving radio waves emitted from the
GPS satellites, we set a constraint on the strain amplitude of the
gravitational wave background $h_{\rm c}$.  We find that the GPS has
already placed a meaningful constraint, and the constraint on the
continuous component of gravitational waves is given as $h_{\rm
c}<4.8\times 10^{-12}(1\slash f)$ at $10^{-2}\lesssim f \lesssim 10^{0}$ Hz, for 
stabilized oscillators with ${\Delta \nu}\slash{\nu}\simeq 10^{-12}$.  Thanks
to the advantage of the Doppler tracking method, seismic oscillations do
not affect the current constraint.
Constraints on $h_c$ in the same frequency range from the velocity measurements by the lunar explorers in the Apollo mission are also derived.

\end{abstract}

\maketitle

\section{Introduction}
The existence of gravitational waves is predicted in general relativity
and many modified gravity theories.  By observing gravitational waves (GWs),
we can not only test the general relativity and modified gravity
theories in strong gravitational fields, but also reveal the merger
process of compact objects through the history of the universe and
physics in the early universe such as inflation ({\it e.g.}
\cite{2011PhRvL.107u1301G}).  In particular, GWs which
originate from the cosmological inflation have almost scale-invariant
spectrum and propagate freely since their generation, and thus the
detection of such scale-invariant GWs can be
considered as a direct proof of the existence of cosmological inflation.
In order to prove the primordial GWs to be
scale-invariant, we have to observe GWs with a wide frequency range 
and high sensitivity.

Precise measurements of primordial GWs at high
frequencies will tell us about the thermal history of the early
universe, which could not be reached in other ways. For example, it has
been suggested that the amplitude of GWs at the
frequency range $f \gtrsim 10^{-2.5}$ Hz can be used to infer the reheating
temperature \cite{2008PhRvD..77l4001N}, and the epoch of quark-gluon
phase transition and neutrino decoupling at lower frequency range 
$10^{-10} \lesssim f \lesssim 10^{-8}$ Hz \cite{2009PhRvD..79j3501K}. 
If the universe underwent a strongly
decelerating phase after inflation realized in quintessential inflation
models \cite{1999PhRvD..59f3505P},
inflationary GWs have a blue spectrum at higher frequency $f\gtrsim 10^{-2}$ Hz
\cite{1999CQGra..16.2905G,2004CQGra..21.1761T}. Therefore, while the
first detection of primordial GWs is expected through
the CMB experiments which probe the waves at smallest frequencies
({\it e.g.} \cite{2000cils.book.....L})
it is significant to explore gravitational
waves with a wide range of frequency.

In addition, measurements of GWs at high frequency will
open a new window on the black hole merger events in the universe.
Matsubayashi {\it et al.} showed that the frequency of gravitational
waves that originate from mergers of intermediate mass black holes
(IMBHs) falls inside the range of $10^{-5}\lesssim f \lesssim 10^{1}$ Hz, and one
can discover or set constraints on these merger events within several
gigapersec from the Earth in future gravitational wave observation
projects \cite{2004ApJ...614..864M}.

In order to observe GWs, many kinds of gravitational wave
detectors have been proposed and developed.  The Laser Interferometer
Gravitational Wave Observatory (LIGO) has already set a strong
constraint on the strain amplitude $h_{\rm c}<4\times 10^{-24}$ at the
frequency of GWs around $10^{2}$ Hz
\cite{2009RPPh...72g6901A,2013PrPNP..68....1R}.  However, seismic
oscillations interfere ground-based observations of gravitational
waves at lower frequencies as $f\lesssim 10$ Hz ({\it see}
\cite{2001PhyU...44R...1G}).  Aiming at this frequency range, Torsion
bar detectors such as TOBA have been developed. Ishidoshiro {\it et al.}
set a constraint on the strain amplitude of the continuous component of
GWs as $h_{\rm c}< 2\times 10^{-9}$ at $f\sim 0.2$Hz
\cite{2011PhRvL.106p1101I}.

At slightly lower frequency range, with some planetary explorers,
constraints such as $h_{\rm c}<1\times 10^{-15}$ at $f=10^{-2}$ Hz \cite{1995A&A...296...13B}
and $h_{\rm c}<2\times 10^{-15}$ at $f=3\times 10^{-4}$ Hz \cite{2003ApJ...599..806A}
have been set by the Doppler tracking method.
However it is difficult to distinguish
signals of GWs as the frequency of targeting
GWs increases higher than $10^{-2}$ Hz due to the noise
in the electric circuits of which are installed in the receiver in this method. 

In this work, we consider the modulation of frequency of radio
waves from GPS satellites by stationary/prompt GWs.
GPS satellites emit precious, stable and high intensity radio waves 
in order to provide positional accuracy for the GPS in navigation. 
It is possible because the oscillators of GPS satellites are designed to 
be synchronized with atomic clocks which are loaded with the satellites.
These oscillators are so stable that 
the fractional error of the frequency of their emitting radio waves 
is suppressed to $\Delta \nu \slash \nu<10^{-15}$
\cite{2008.gps.standard}.

The GPS method is an application of the Doppler tracking method which has two
advantages. First, GPS method can probe in the frequency range 
$f\gtrsim 10^{-2}$ Hz where the conventional Doppler tracking methods can not
reach because GPS satellites are much closer than the planetary
explorers, while the distance is large enough for the noise due to
seismic oscillations to be negligible as shown in the next section.
Secondly, the radio waves from GPS 
satellites can be detected everywhere and everytime on the ground.
This condition will be suitable to make cross correlation study and
crucial to detect GWs from prompt events. 

In this study, we set a constraint on the amplitude of GWs
for $0.01 \le f \le 1$ Hz by using the Doppler tracking method with
GPS disciplined oscillators (GPSDOs). The GPSDO is the stable oscillator with
a quartz oscillator whose output is controlled to agree with the signals
from GPS satellites \cite{2008.gps.standard,2005ptti.conf..677L}.  To
have GPSDOs operating with high stability, the amplitude of the
continuous components of GWs should be small.  Recently,
the frequency stability of GPSDOs for a short time interval about from
seconds to a few hundred seconds has been reached to the level of
${\Delta \nu}\slash{\nu} \simeq 10^{-12}$ \cite{2008.gps.standard,2005ptti.conf..677L}. 
This short time stability of the
frequency of GPSDOs enables us to set a constraint on the strain
amplitude of the continuous GWs.

This paper is organized as follows. We estimate the effects of
GWs on the 
measurements of the radio waves from GPS satellites and derive the 
constraint in \S 2.  In our
formulation, we adopt TT gauge to describe the GWs.  In
\S 3, we discuss implications of the result to the
merger events of IMBHs and inflationary GWs. We conclude
this paper in \S 4. Throughout 
this paper, $\nu$ and $f$ mean the frequency of the electro-magnetic
waves and the GWs, respectively. The speed of light is
denoted by $c$.  A dot represents a partial derivative respective to the
physical time, {\it i.e.} $\dot{x}\equiv {\partial x}\slash{\partial
t}$.

\section{Effects of GWs on GPS measurements}
In this section, we consider the effect of gravitational wave
backgrounds (GWBs) on the radio waves emitted by GPS satellites.  Here
we assume that GWBs are expected to be isotropic and stationary ({\it
see.} \cite{2000PhR...331..283M}).  For GWs whose
wavelength are longer than the typical distance between GPS satellites
and detectors on the ground, the frequency of radio waves emitted by the
satellites is modulated as 
\begin{equation} 
\dfrac{\Delta \nu}{\nu}=
\dfrac{l}{2c}\dot{h}(t)\sin^{2}\theta ~,\label{fractional01}
\end{equation}
where $\dot{h}(t)$, $\theta $ and $l$ are the time derivative of the
amplitude of GWs at time $t$, the angle between the 
directions of propagation of radio waves and gravitational
waves, and the distance between the GPS satellite and the observer,
respectively.

By assuming GWs are monotonic and plane waves
parallel to the $z$-axis, the amplitude of GWs $h(t)$ and its time derivative $\dot{h}(t)$
can be written with the strain $h_{\rm c}$ as
\begin{eqnarray} 
h(t)&=& h_{\rm c}\sin \left(2\pi f \left(t-\dfrac{z}{c}\right)+\phi \right)~,\\
\dot{h}(t)&=& 2\pi f h_{\rm c}\cos \left(2\pi f \left(t-\dfrac{z}{c}\right)+\phi \right)~,\label{differential}
\end{eqnarray}
where $\phi $ is the phase of GWs at $z=t=0$.

From Eq. (\ref{fractional01}), signals of GWs generate
an additional shift of frequency of radio waves. When one considers the
case where $\theta =\pi \slash 2$, the effect of GWs on
the frequency modulation of electro-magnetic waves can be estimated as
\begin{equation} 
\dfrac{\Delta \nu}{\nu}=
\dfrac{\pi fl}{c}h_{\rm c}~.
\label{fractional02}
\end{equation}
In reality, the signal sourced by GWs is buried within 
the noise. Therefore, if one can receive 
the radio waves which is emitted at a distance of $l$ 
with a time variance of the frequency fluctuations $\sigma$
\cite{baran2010modeling},  
one can set the upper bound of the strain amplitude of gravitational
waves as, 
\begin{equation} 
h_{\rm c}<\dfrac{c}{\pi lf}\sigma ~.\label{hc}
\end{equation}

For the GPS, the distance $l$ is approximately $2\times 10^{7}$ m and 
the standard variation of frequency from GPS satellites converges to 
$\sigma \simeq 1\times 10^{-12}$ 
by integrating the signal from GPS satellites for the period 
from one second to one hundred seconds 
\cite{2005ptti.conf..677L,2008.gps.standard}.
By receiving the signal from the GPS for a period $t_{\rm i}$, 
the frequency range of the GWs which one can probe is limited
as $f\ge t_{\rm i}^{-1}$. By combining it with the condition  
that the wavelength of GWs is longer than 
the distance between the GPS satellites and the observer, 
the probed range of frequency with the GPS can be written as
\begin{equation} 
t_{\rm i}^{-1}\le f \le \dfrac{l}{c}~.
\end{equation}

\begin{figure}
 \begin{center}
  \includegraphics[width=85mm]{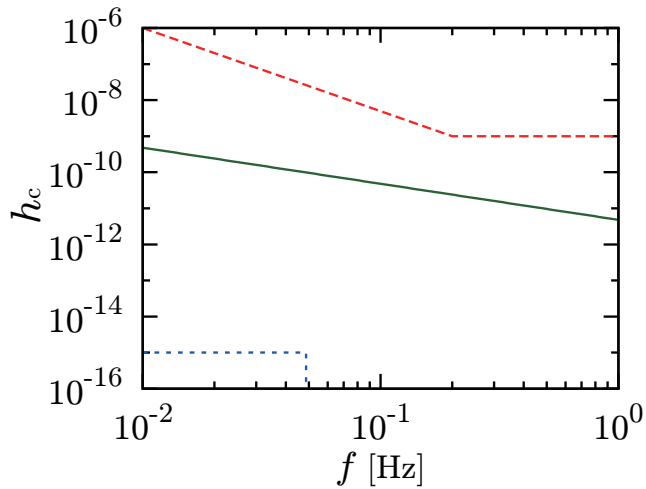}
 \end{center}
  \caption{The upper limit on the strain amplitude of GWBs from GPS
 satellites (solid line).  The dashed line represents the constraint
 from the torsion bar detector \cite{2011PhRvL.106p1101I}.  The dotted
 line represents the constraint from the Doppler tracking
 method\cite{1995A&A...296...13B}. }  \label{fig.1}
\end{figure}

From Eq. (\ref{fractional01}) 
because the amplitude of
the modulation is proportional to the frequency of GWs
$f$, the strain amplitude $h_{\rm c}$ is constrained tighter for higher
frequency.  By substituting the numbers into Eq. (\ref{hc}), we can set
a constraint on the strain of GWBs as
\begin{equation} 
h_{\rm c}<4.8\times 10^{-12}\left(\dfrac{1 {\rm Hz}}{f} \right)
~.\label{constraintGPS}
\end{equation}
In figure \ref{fig.1}, we plot the constraint on the strain amplitude of
the continuous component of GWs and compare the result
from the torsion bar detector \cite{2011PhRvL.106p1101I}.  We find that
the GPS gives a tighter constraint on the relevant frequency range.

The GPS method does not suffer seismic oscillations that disturb
ground-based observations. The reason can be given as follows.
In the GPS method, the effect of GWs can be seen 
as changes of observed distances between the satellites and observers.
When plane and monotonic GWs come to an observer, 
the observed distance $l$ between the satellite and the observer on the
ground can be written in terms of the proper distance $l_0$ as 
\begin{equation} 
l=l_{0}\left(1+h \right)+x_{\rm s}~,\label{equivalentEq}
\end{equation}
where $h$ is the amplitude of GWs and $x_{\rm s}$ is the
amplitude of seismic oscillations.  Thus the effect of seismic
oscillations relative to the strain amplitude of GWs can
be characterized by $x_{\rm s}\slash l_{0}$.  Shoemaker {\it et al.}
reported a typical power spectrum of seismic oscillations as
\cite{1988PhRvD..38..423S}
\begin{equation} 
x_{\rm s}\simeq 3\times 10^{-7} \left(f\slash 1~{\rm Hz}
				\right)^{-2}~{\rm [m]}~. 
\end{equation}
Because $l_{0}\simeq 2.0\times 10^{7}$ m, the effect of seismic
oscillations is suppressed as,
\begin{equation} 
\dfrac{x_{\rm s}}{l_{0}}\simeq 
1.5\times 10^{-14}\left(f\slash 1~{\rm Hz}
				\right)^{-2}~.
\end{equation}
This is smaller than the upper limit of our constraint, Eq
(\ref{constraintGPS}). 

In addition, because GPS satellites fly much closer to the ground
than the planetary explorers which have been used in the Doppler
tracking method, we can set a constraint on the strain amplitude of GWBs
in the higher frequency region than those of ULYSSES and Cassini (
\cite{1995A&A...296...13B,2003ApJ...599..806A}).

\begin{figure}
 \begin{center}
  \includegraphics[width=85mm]{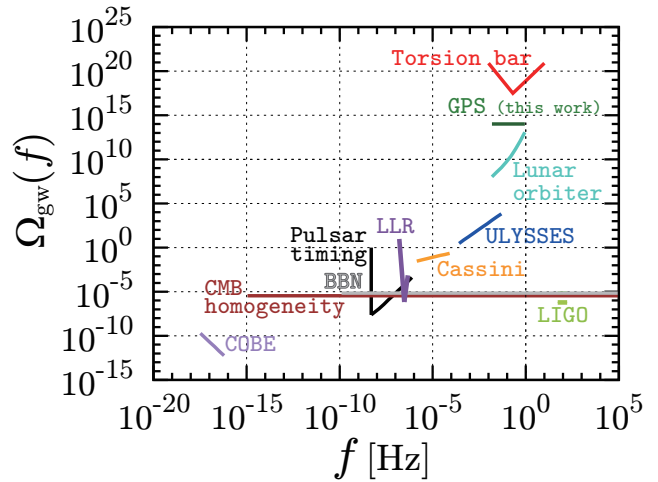}
 \end{center}
  \caption{Summary of the constraints on GW background in terms of
$\Omega_{\rm gw}(f)$, which includes  {\tt COBE} \cite{2000PhR...331..283M}, {\tt CMB
homogeneity} and {\tt BBN} \cite{2006PhRvL..97b1301S}, {\tt Pulsar
timing} \cite{2006ApJ...653.1571J}, {\tt LLR}
\cite{1981ApJ...246..569M,2009IJMPD..18.1129W}, {\tt Cassini}
\cite{2003ApJ...599..806A}, {\tt ULYSSES} \cite{1995A&A...296...13B},
{\tt Lunar orbiter} \cite{2001Icar..150....1K},
{\tt Torsion bar} \cite{2011PhRvL.106p1101I}, {\tt LIGO}
\cite{2009RPPh...72g6901A,2013PrPNP..68....1R}, and {\tt GPS}.}  \label{fig.2}
\end{figure}

The intensity of GWBs can be characterized by the dimensionless
cosmological density parameter $\Omega_{\rm gw}(f)$.  The parameter is
defined as
\begin{equation} 
\Omega_{\rm gw}(f)=\dfrac{10\pi^{2}}{3H_{0}^{2}}(fh_{\rm c})^{2}~,
\end{equation}
where $H_{0}$ is the Hubble constant, and the Planck collaboration
reported as $H_{0}=67.11~{\rm km/sec/Mpc}=2.208\times 10^{-18}~{\rm
sec}^{-1}$\cite{2013arXiv1303.5076P}.  Then the constraint Eq.~(\ref{constraintGPS}) can be written as
\begin{equation} 
\Omega_{\rm gw}(f)<1.7\times 10^{14} ~~{\rm for}~10^{-2} \lesssim f \lesssim 10^{0}[{\rm Hz}]~.
\end{equation}
In figure \ref{fig.2}, we compare our constraint with the previous ones.
It can be seen that the Doppler tracking method with the GPS is setting
a constraint on the amplitude of continuous component of gravitational
waves in the frequency range of $10^{-2} \lesssim f \lesssim 10^{0}$ Hz, a window between
the constraints from the Torsion bar and the planetary explores.

\section{Discussion}

In this work, we show that satellites with atomic clocks are available
for setting constraints on the strain amplitude of GWs
at $10^{-2}\lesssim f \lesssim 10^{0}$ Hz.  The constraints based on the Doppler tracking
method with planetary explorer such as ULYSSES and Cassini are mainly
limited from the stability of the hydrogen maser clock on the ground
$\sim 10^{-15}$ at the frequency range $10^{-6}\lesssim f\lesssim 10^{-2}$ Hz. 
The stability of the optical lattice clock is expected
to reach $10^{-18}$ \cite{2003PhRvL..91q3005K}.  If future explorers 
have the optical lattice clock on board, they will become useful
instruments for detection/setting constraints on the strain amplitude
of GWs ({\it see also }\cite{2006LRR.....9....1A}).

Determination accuracy of frequency fluctuations of received radio waves 
from GPS satellites depends mostly on the time resolution of 
the received electric-signal in the A\slash D converter
of the GPS receiver $\sigma_{\rm r}$. It is reported that $\sigma_{\rm r}
\ge 2\times 10^{-13}\slash t_{\rm i}$ and the largest error comes from
the frequency transfer \cite{2005ptti.conf..149F}.  It is difficult to
improve the GPS constraints on the strain amplitude until the 
resolution of the quantization in the A\slash D converter is much improved.

Here let us apply our result to setting a constraint on the
merger events of compact objects. In particular, it is predicted that
mergers of the binary of IMBHs emit large GWs in the
relevant frequency range, while it is difficult to detect the events
through X-rays or radio waves if the systems do not have accretion
disks.  Therefore observations or constraints of GWs
enable one to estimate the number density of the binary of IMBHs and the
merger events.  

During the quasi-normal (QNM) phase of the IMBHs mergers, in which
merging IMBHs are expected to emit the largest amplitude of 
GWs, the amplitude $h_{\rm QNM}$, typical
frequency $f_{\rm QNM}$ and the duration time of the event $t_{\rm QNM}$
are given by \cite{2004PThPS.155..415M},
\begin{eqnarray} 
f_{\rm QNM}&\simeq &4\times 10^{-2}\left(\dfrac{M}{10^{6} M_{\odot}} \right)^{-1}~[{\rm Hz}],\label{f.insp}\\
h_{\rm QNM}&\simeq &2\times 10^{-12}\left(\dfrac{M}{10^{6} M_{\odot}} \right)
\left(\dfrac{\varepsilon }{10^{-2}} \right)^{\frac{1}{2}}
\left(\dfrac{R}{4 {\rm kpc}}\right)^{-1}
~,\label{h.insp} \\
t_{\rm QNM}&\simeq &30\times 
\left(\dfrac{M}{10^{6} M_{\odot}} \right)~[{\rm sec}].\label{t.insp}
\end{eqnarray}
Here $\varepsilon$ is the eccentricity of the orbit, $R$ is the distance
to the binary, and we assumed that two IMBHs have the same mass $M$.
From Eqs. (\ref{constraintGPS}), (\ref{f.insp}) and (\ref{h.insp}), 
one can rule out merger events of IMBHs from the GPS constraint as
\begin{eqnarray} 
R&\gtrsim&0.1\left(\dfrac{\varepsilon }{0.01} \right)^{-1\slash 2} [{\rm kpc}]~\\
& &{\rm for}~4\times 10^{4}M_{\odot} \le M \le 4\times 10^{6}M_{\odot}~. \notag
\end{eqnarray}
The minimum and maximum masses are 
determined from the frequency range we can probe in this method, i.e. Eq.(4).

Frequency modulation signals induced by GWs may be
disturbed by the plasma effect in the ionosphere and the atmosphere. 
Fluctuations of the column density of free electrons in the plasma,
called the dispersion measure (DM), induce those in frequency of GPS
radio waves as
({\it e.g.} \cite{2004tra..book.....R,Jackson})
\begin{equation} 
\dfrac{\Delta \nu}{\nu}=\dfrac{e^{2}\nu^{-2}}{2\pi m_{\rm e}cr}{\rm DM}~,
\end{equation}
where ${\rm DM}=\int_{0}^{r}n_{\rm e}ds$ is the column density of
electrons, $m_{\rm e}$ and $e$ are the mass and the electric charge of
an electron, respectively, and $r$ is the distance between the GPS
satellite and the observer.  By comparing the above equation with
Eq. (\ref{fractional02}), one can see that the frequency dependences are
different between the modulations induced by GWs and the
plasma effect.  Therefore the modulation originated from the plasma can
in principle be removed by using multiple frequencies.  Some of the
GPSDOs observe the two bands of GPS radio waves (L1 \& L2) \footnote{The
frequency of the band L1 and L2 are 1.57542 GHz and 1.2276 GHz,
respectively.}  and take into account this modulation.

However, even when GPSDOs give the best performance, the distance at
which one can detect the merger of IMBHs with GPSDOs is limited only to
0.1 kilopersec from the Earth.  In order to detect mergers of IMBHs and compact
objects and the GW background from inflation with realistic amplitude,
space-based gravitational wave detectors are needed. 
As future gravitational wave detectors, Laser Interferometer Space Antenna (LISA)
and Deci-hertz Interferometer Gravitational wave Observatory (DECIGO)
are in progress.  LISA and DECIGO are expected to reach $h_{\rm c}\simeq 3\times
10^{-21}$ at $f\simeq 6 \times 10^{-3}$ Hz and $h_{\rm c}\simeq 2\times 10^{-24}$ at
$f\simeq 0.3$ Hz, respectively
\cite{2013arXiv1305.5720C,2009JPhCS.154a2040S}.  These sensitivities
will enable us to detect almost all the merger events of IMBHs with mass
$M\sim 10^{3}M_{\odot}$ within the current horizon \footnote{The current
horizon scale is approximately equal to the distance to the last
scattering surface of CMB $d_{\rm A(CMB)}$. Jarosik {\it et al.}
reports that $d_{\rm A}=14.116^{+0.160}_{-0.163}$ Gpc
\cite{2011ApJS..192...14J}.}, and to reveal the properties of the strong
gravity field and the cosmological inflation
\cite{2004PThPS.155..415M}.

 Finally let us discuss another constraint that can be obtained from the lunar
orbitting explorers in a similar way to the GPS constraint. 
In the Apollo mission, in order
to study the gravity field of the moon,
lunar explorers such as Apollo 15 and 16 measured 
the change in distance between the
explorers and the Earth precisely 
via S-band transponders \cite{2001Icar..150....1K}.
The typical distance between the explorers and the ground is $l\approx
3.8\times 10^{8}$ m, which is much longer than that of the GPS case.
It was reported that anomalous oscillating motions had never  been
found  in their every ten-second sampling data with $1\times
10^{-4}~$m/sec accuracy \cite{2001Icar..150....1K}.
From Eqs. \eqref{differential} and \eqref{equivalentEq}, this 
result is translated to
the strain amplitude of GWs as $h_{\rm c}<2.6\times 10^{-13}$ at
the frequency range $f < c/l \simeq 0.8$Hz \footnote{ To be more
precise, the distance between the moon and the earth is so long that
this constraint should be corrected at frequencies above 0.1Hz as
$h_{\rm c}<2.6\times 10^{-13}\sqrt{1+\left(f/0.16~{\rm
[Hz]}\right)^{2}}$.
At $f=1$ Hz, we obtain the constraint $h_{\rm c}<1.6\times 10^{-12}$.
}.
The constraint is also depicted in Fig. \ref{fig.2}.
Recent lunar explorers such as Kaguya \cite{2009Sci...323..900N} 
and GRAIL \cite{2013Sci...339..668Z,2010EGUGA..1213921Z} may improve this upper bound. 
In addition, future
measurements with lunar surface transponders 
such as those proposed by Gregnanin {\it et al.} 
\cite{2012P&SS...74..194G} will be available for setting 
constraints on $h_{\rm c}$.

\section{Conclusion}
We set a constraint on the strain amplitude of the continuous component
of GWs as $h_{\rm c }<4.8\times 10^{-12}\left(1[{\rm
Hz}]\slash f \right)$ at the frequency range $10^{-2}\lesssim f \lesssim 1$ Hz 
with the radio waves emitted by GPS
satellites in operation. 
Because the distance between GPS satellites and observers
is order $10^{7}$ m, seismic oscillations do not affect the
constraints on the strain amplitude.  The sensitivity to the
GWs is limited to that of the A\slash D converter on the
GPS receiver at the frequency range.

\section{Acknowledgments}
{
We thank the referee, Peter L. Bender, for making us aware of the fact that past and future lunar orbitar experiments can be used to set better constraints on the amplitude of continuous GWs. Our thanks also go to Jun'ichi Miyao for discussion on the characteristics of the GPS, and Seiji Kawamura, Yoichi Aso on the noise in the electric circuits installed on the gravitational wave detectors.}
This work is supported in part by scientific research expense for Research Fellow of the Japan Society for the Promotion of Science from JSPS Nos. 24009838
(SA) and 24340048 (KI).

\bibliography{reference2013b}
\end{document}